\documentclass[american,english]{aastex}
\usepackage[T1]{fontenc}
\usepackage[latin9]{inputenc}
\usepackage{graphicx}
\usepackage{subscript}

\makeatletter

\providecommand{\tabularnewline}{\\}

\slugcomment{}
\shorttitle{}
\shortauthors{}

\usepackage{babel}

\makeatother

\usepackage{babel}
\begin{document}

\title{The Extreme Optical Variability of J0948+0022}

\author{Jeremy D. Maune, H. Richard Miller, Joesph R. Eggen}

\affil{Department of Physics and Astronomy, Georgia State University, Atlanta,
GA 30303-3083}

\email{maune@chara.gsu.edu}
\begin{abstract}
We report on observations of the optical variability of the radio-loud,
narrow-line Seyfert 1 galaxy J0948+0022 on time scales ranging from
minutes to years. Implications regarding recent suggestions that the
object may constitute a prototype for an emerging class of blazar-like
objects similar to FSRQs are discussed. The optical microvariability
observed for J0948+0022 is found to be similar to that found for a
typical LBL blazar. Based on observations of J0948+0022 in a flaring
state and a significantly lower state, one can demonstrate that these
rapid variations are most likely originating in the relativistic jet
and not in the accretion disk. 
\end{abstract}


\section{Introduction}

Blazars are a class of AGN that is viewed with the relativistic jet
oriented near the line of sight to the observer. Observationally,
they demonstrate a classic double-peaked spectral energy distribution
(SED) and are detectible across the electromagnetic spectrum, notably
existing in both the radio and gamma ray regimes; in some cases, these
objects are even detected at energies in the TeV range. Blazars are
also highly variable at all wavelengths and on all observed time scales,
from minutes to years. In addition, the optical emission of blazars
demonstrates significant polarization that is variable in both magnitude
and position angle.

In the past few years, groups such as Abdo et al. (\citealt{Abdo1})
and Yuan et al. (\citealt{Yuan1}) have demonstrated that a subset
of Seyfert galaxies \textemdash{} i.e., very radio-loud (R>50) narrow-line
Seyfert 1s (VRL-NLSy1s) \textemdash{} appear to share many of the
same observational traits as blazars. They have suggested that these
objects constitute a new subclass of blazar. This would be of particular
interest since previously blazars have exclusively been found to be
hosted by elliptical galaxies, whereas Seyferts are typically found
in spirals.

The prototype for this potential new class is the VRL-NLSy1 J0948+0022.
It has already been shown that J0948+0022 is associated with a variable
gamma ray source and that it features a double-peaked SED similar
to that of a typical blazar (\citealt{Foschini1}). The object is
also known to be radio loud, with a radio loudness parameter R>log(2.5)
(\citealt{Yuan1}). The redshift (z=0.584) obtained by the Sloan Digital
Sky Survey (SDSS) confirms that it is an extragalactic source (\citealt{SDSS}).
However, although violent optical variability remains one of the hallmark
characteristics of blazar activity, the optical emission of the object
has yet to be thoroughly studied. This is likely due to the unfortunate
faintness of the object. The authors have observed J0948+0022 to fade
to a minimum of m\textsubscript{R}= 20 during a recent, particularly
faint state.

In this paper, we report on a recent monitoring program initiated
to study the optical flux of J0948+0022 in an attempt to further investigate
the nature of its variability and determine if it exhibits behavior
on blazar-like time scales. It should also be noted that a forthcoming
companion paper (\citealt{Eggen1}) will detail the optical polarized
emission and gamma ray emission from the object over a similar time
period.

\section{Observations}

Data for this study were primarily obtained using the 72-inch Perkins,
42-inch Hall, and the 31-inch NURO telescopes at Lowell observatory
in Flagstaff, Arizona. Additional observations were acquired from
the 1.3 meter telescope at the Cerro Tololo Interamerican Observatory
in Chile through the SMARTS consortium. Specifications for these instruments
can be found in Table \ref{tab:Inst-Specs}. Data from the Lowell
telescopes were used to monitor intra-night to nightly trends, whereas
the SMARTS telescope was used to monitor longer term behavior. Images
were obtained exclusively in the standard Johnson R band. All images
taken at Lowell observatory were reduced in the standard fashion.
Every evening in which observations were acquired, multiple bias frames
were obtained and combined into a master calibration file. Flat fielding
images were taken at least once per observing run using an in-dome
screen. These were then combined into a single master flat. The calibration
files were then applied to all science images acquired during the
observing session. Since the detectors are cooled significantly below
ambient temperature, no dark frames were required. Images obtained
from SMARTS were processed in a similar manner, except that the actual
flat fielding and bias subtraction was done on-site by the observer
in Chile to be forwarded pre-reduced to the authors.

In order to generate accurate light curves for the object and to directly
compare data from different telescopes, several in-field comparison
stars were selected for use in differential photometry. The object
was then observed on a photometric night along with several fields
containing stars of known magnitude taken from the Landolt list of
equatorial stars (\citealt{Landolt1}). Images of J0948+0022 and the
Landolt stars were then observed at as close to the same air mass
as possible to correct for atmospheric effects. Using the known values
of the Landolt objects, the true apparent (non-instrumental) magnitudes
of the selected stars in the field of J0948+0022 could then be derived.
The stars thus chosen are indicated in the finding chart in Figure
\ref{fig:Finding_Chart}.

The in-field standards were confirmed to be stellar sources using
the on-line tools of the eighth data release of the SDSS. However,
it should be noted that the stars were selected based entirely on
the fact that they were relatively close in brightness to J0948+0022.
In order to ensure that these stars are truly non-varying sources,
each star was treated as an unknown object and its magnitude was derived
using the other eight comparison stars in a process identical to that
used to find the magnitude of J0948+0022. This was done for every
image of the field taken by the three Lowell telescopes across the
observing session; the SMARTS telescope was excluded from this analysis
because it has a much smaller field of view and it was impossible
to observe all nine stars on the same image. The resulting light curves
confirm that the selected standard stars are stable. The individual
R magnitudes and uncertainties of these stars are detailed in Table
\ref{tab:Check_Star_Info}.

After the finding chart depicted in Figure \ref{fig:Finding_Chart}
had been constructed, the authors became aware of another chart for
the same field constructed by the Landessternwarte Königstuhl (LSW)
at the University of Heidelberg, Germany. By chance, all of the stars
chosen to serve as comparison objects on the LSW chart also appear
in Figure \ref{fig:Finding_Chart}. However, the two finding charts
disagree as to the magnitudes of these stars, an issue that casts
doubt onto any published results using either source. An attempt to
reconcile the differences between them is detailed in the appendix
of this paper.

\section{Results}

The long-term optical behavior of J0948+0022 over the past two years
is shown in Figure \ref{fig:monthly_lc}. Uncertainties for the observations
shown in this figure are typically between 0.01 to 0.05 magnitudes.
For purposes of comparison, the sixth standard star has been plotted
on the 2012 light curve as well; the horizontal lines bounding these
data indicate the range in magnitude that would be expected of the
star given the observational uncertainty. In all locations where the
star strays from these boundaries, it was found that the other stars
also showed atypical behavior, indicating that these events are likely
due to atmospheric effects rather than intrinsic variability. The
check star demonstrates comparable behavior in the 2011 light curve,
but this is not shown due to the fact that the check star data would
have overlapped with the object data on the plot.

J0948+0022 shows extensive variability with a total observed range
of 2.34 magnitudes, with a maximum brightness of m\textsubscript{R}
= 17.69 occurring in May of 2011 and a minimum of m\textsubscript{R}
= 20.03 in April of 2012. Remarkably, only a month after this minimum,
the object had flared to nearly its previously observed peak intensity,
reaching m\textsubscript{R} = 17.87. On shorter time scales, the
object demonstrates inarguable microvariable behavior. Figure \ref{fig:weekly_lc}
shows observations obtained over six nights at the end of March 2011,
while Figure \ref{fig:hourly_lc} details the same six nights on an
individual basis. These nights represent the highest time resolution
photometry reported to date for J0948+0022.

The optical variability behavior demonstrated by J0948+0022 is not
only comparable to that of blazars both in amplitude (\citealt{Noble1})
and in observed structure (\citealt{Miller2 and Noble2}), but would
be remarkable even by a blazar's high standards. Clear trends well
above the noise level of the standard stars can be seen in four of
the six nights shown in Figure \ref{fig:hourly_lc}. On March 27,
there is a slow fading of the object by $\sim0.15$ mag observed over
5 hours without any statistically significant short term variations
superposed on this decline. On March 28, a major event is observed
near $20110328.2$ when J0928+0022 rapidly flares $\sim0.3$ mag in
$\sim1$ hour. Following this event, a decline of $\sim0.15$ mag,
with short-term flickering of $sin0.1$ mag superposed, is observed
over the next four hours. On March 29, J0948+0022 shows no evidence
of significant variations as it resides near R $\sim19.0$ mag. On
March 30, J0948+0022 exhibited two significant microvariability events
with amplitudes of $\sim0.15$ mag and durations of $\sim2$ hours.
On March 31, the observed variations are less well-defined due to
larger errors, but there are clearly variations present with an amplitude
of $\sim0.3-0.4$ mag during the night. On the night of April 1, truly
spectacular variability is observed with a total amplitude of $\sim0.9$
mag during $\sim4$ hours. J0948+0022 brightened and exhibited two
significant and separate events separated by $\sim2.5$ hours! At
a rate of change of $0.2-0.3$ mag/hour, this is one of the most rapid
events observed for any blazar-like object. 

The frequent, large-amplitude microvariability observed for J0948+0022
is consistent with the variability behavior reported for low energy
peaked BL Lac (LBL) blazars (\citealt{Miller2 and Noble2}), a class
that includes flat spectrum radio quasars (FSRQs). Thus, these results
support the suggestion that J0948+0022 is more similar to FSRQ-type
blazars than their high energy peaked counterparts, which are not
observed to exhibit very short time scale variability at such high
amplitudes.

\section{Conclusions}

Although microvariability is one of their defining characteristics,
not even the most active blazar demonstrates microvariability every
night it is observed. In fact, on rare occasions even radio quiet
AGN will undergo rapid optical variations within a single observing
session. A more complete method of comparing the relative activity
of VRL-NLSy1s to that found for blazars would be to compare their
duty cycles. A duty cycle is defined as the ratio between the amount
of time an object is observed in a microvarying state to the total
time it was observed. Carini et al. (\citealt{Carini1}) previously
investigated the occurrence of microvariability in radio-quiet Seyferts,
radio-loud Seyferts (where radio loudness was defined as R>10), and
blazars. Although this earlier study suffered from a somewhat limited
sample size, it was found that the likelihood of observing rapid optical
variations within these classes of objects to be 10\%, 19\%, and 45\%
respectively. A similar analysis of the data presented in this paper
reveals that J0948+0022 demonstrates a duty cycle of 57\%, a high
level of activity far more consistent with what would be expected
for a blazar than a normal radio-loud Seyfert galaxy.

Given the existence of this rapid variability, it becomes possible
to explore its origin. The physical cause of the optical microvariability
in blazars has been a matter of discussion since the phenomenon was
first shown to exist. Broadly, two possible scenarios have been put
forth as likely explanations. In the first, microvariability is generated
within the accretion disk of the central black hole through magnetic
flares, hot spots, localized obscuration events, or any other circumstance
that that could lead to a local change in brightness. The observed
variability would then be caused by the combined flux from a variety
of independently varying sources superimposed upon one another, neatly
explaining how the flux can change so rapidly (\citealt{Mangalam1}).
Alternatively, perturbations, shocks, or bends within the relativistic
jet itself could produce rapid variability (\citealt{Marscher1}).

As was shown by Miller et al. (\citealt{Miller1}), it is possible
to distinguish between these two scenarios by observing the object
in both a faint state and a bright state. Since the jet is expected
to dominate the observed flux during outbursts, the relative contribution
from the accretion disk will be minimized when the object is bright
and largest when the object is faint. This means that any variations
originating from the disk will contribute a relatively smaller portion
of the total flux in the bright phase. Therefore, if microvariations
originate in the accretion disk, the amplitude of the microvariability
should be greatest during faint states and minimized during bright
states. If the microvariability is instead caused by events associated
with the jet, whenever the jet undergoes a flaring event those variations
will likewise be boosted. In this case, the fractional amplitude of
the microvariability will be similar in both the high and low states.

Although J0948+0022 did not undergo a major flare event during the
reported observing session, Liu et al. (\citealp{Liu1}) investigated
the behavior of the object in a higher state during April 2009. This
group indicated their observations supported a jet origin for the
microvariability on their own. However, they did not consider the
possibility that the microvariability could have arisen from within
the accretion disk. During the night of April 25th, the Liu group
observed the object to vary between m\textsubscript{R}=16.7 to m\textsubscript{R}=17.3,
with a range of microvariation of 0.5 magnitudes. This activity is
comparable to what is seen in Figure \ref{fig:hourly_lc} of this
paper, in which the object was generally found closer m\textsubscript{R}=
18.5. Given the previous discussion, this suggests that the most likely
origin for the microvariability is within the jet.

In a related matter, given that clear, discrete events can be detected
in the microvariability it should in principle be possible to make
some determination about the size of the emitting region based on
light travel time arguments. Unfortunately, such an exercise would
depend on knowing the value for the Doppler boosting factor for the
object. While attempts to find this value have been made using data
from the VLBA (\citealt{Foschini1}) such studies suffer from high
uncertainties due to the compact nature of the object. Therefore,
such an analysis will have to wait until more precise measurements
can be made.

In summary, there is strong evidence that the characteristics of J0948+0022
are more similar to those of known blazars than typical Seyfert galaxies.
In addition to the previously published double peaked SED, variable
gamma ray emission, and radio footprint, the object has been shown
to be strongly variable in the optical regime on both short and long
time scales. Further, J0948+0022 demonstrates microvariability much
more frequently than that found for other Seyfert galaxies and at
a rate that is comparable to those of blazars. Finally, the nature
of the observed microvariability suggests that it does not originate
in the accretion disk of the central black hole, but rather in the
relativistic jet.

\section{Appendix}

\begin{figure*}[t]
\noindent \centering{}\includegraphics[scale=0.66]{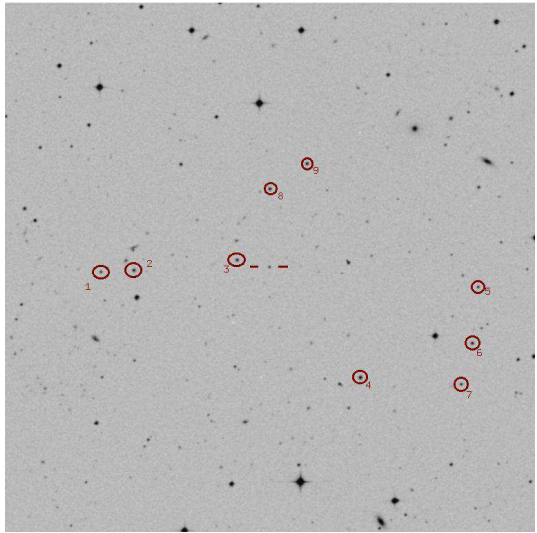}\caption{Finding chart for J0948+0022 with object highlighted. Field of view
is 12.9' x 12.9'\label{fig:Finding_Chart}}
\end{figure*}

\begin{figure*}[t]
\centerline{\includegraphics[scale=0.33,angle=90,origin=c]{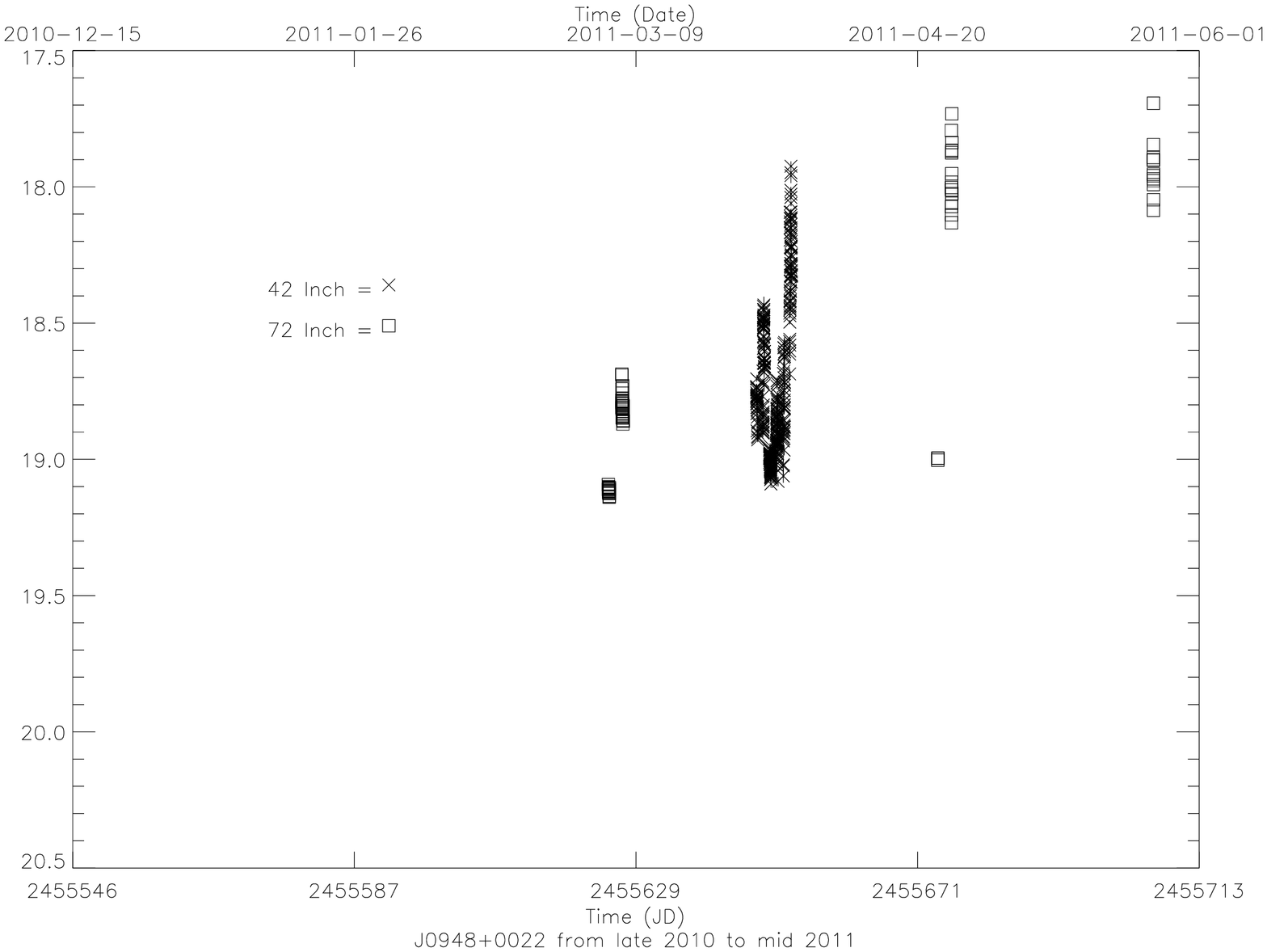}\includegraphics[scale=0.33,angle=90,origin=c]{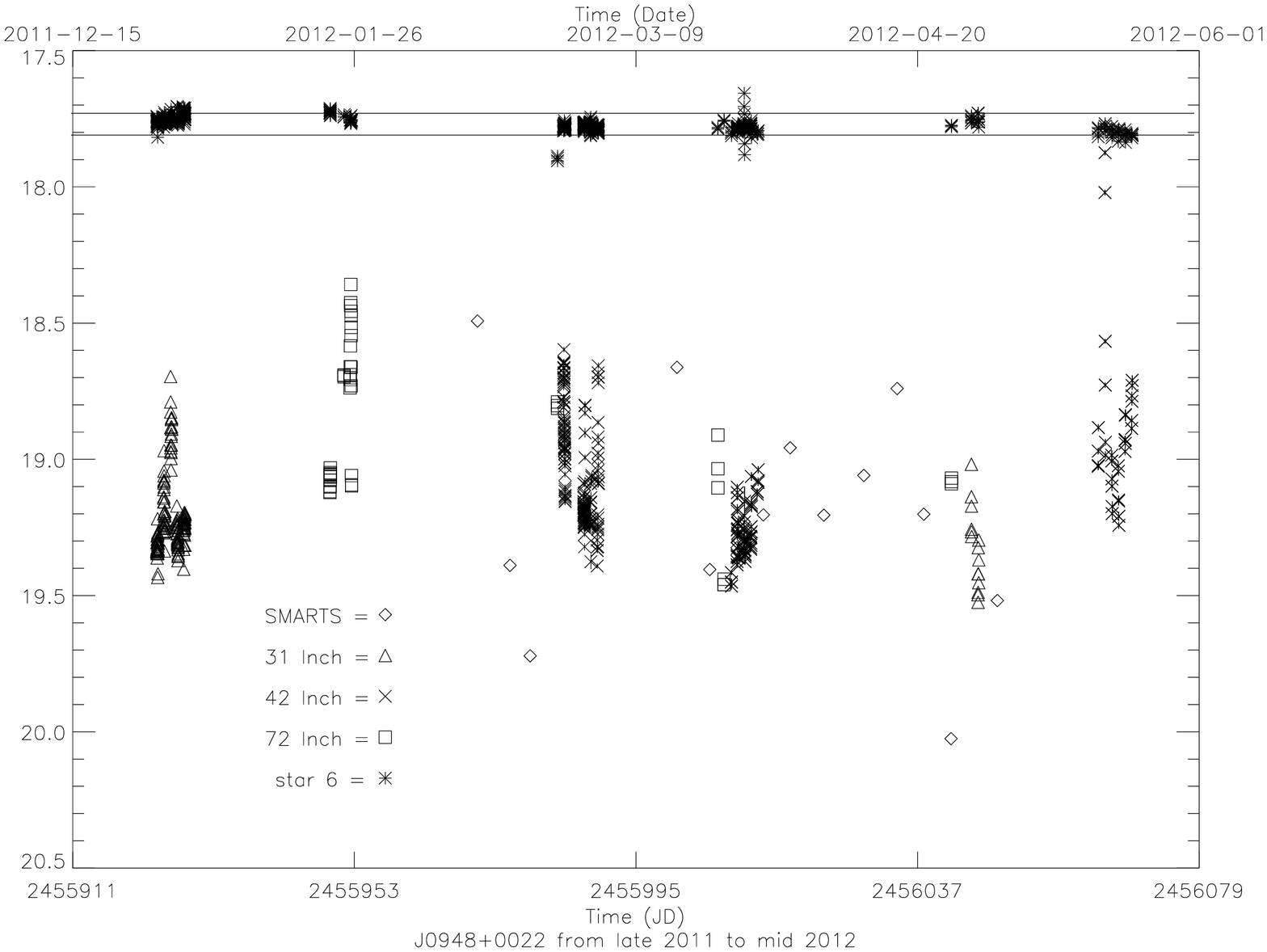}}\caption{Monthly light curves for J0948+0022 for 2011 (left) and 2012 (right).
All images were taken in the R band.\label{fig:monthly_lc}}
\end{figure*}

\begin{figure*}[t]
\centering{}\includegraphics[scale=0.3,angle=90]{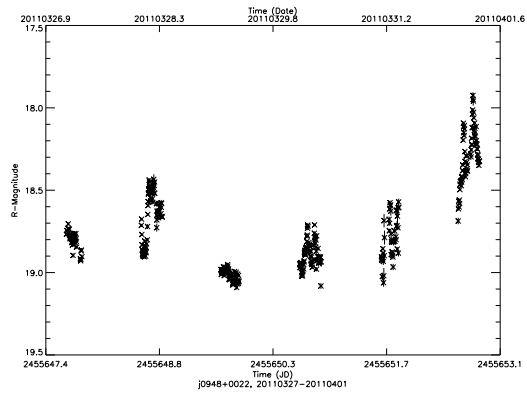}\caption{Behavior of the object on a nightly timescale. Data was obtained exclusively
from the Hall telescope on the nights of March 27 - April 1, 2011.\label{fig:weekly_lc}}
\end{figure*}

\begin{figure*}[p]
\noindent \centerline{\includegraphics[scale=0.3,angle=90]{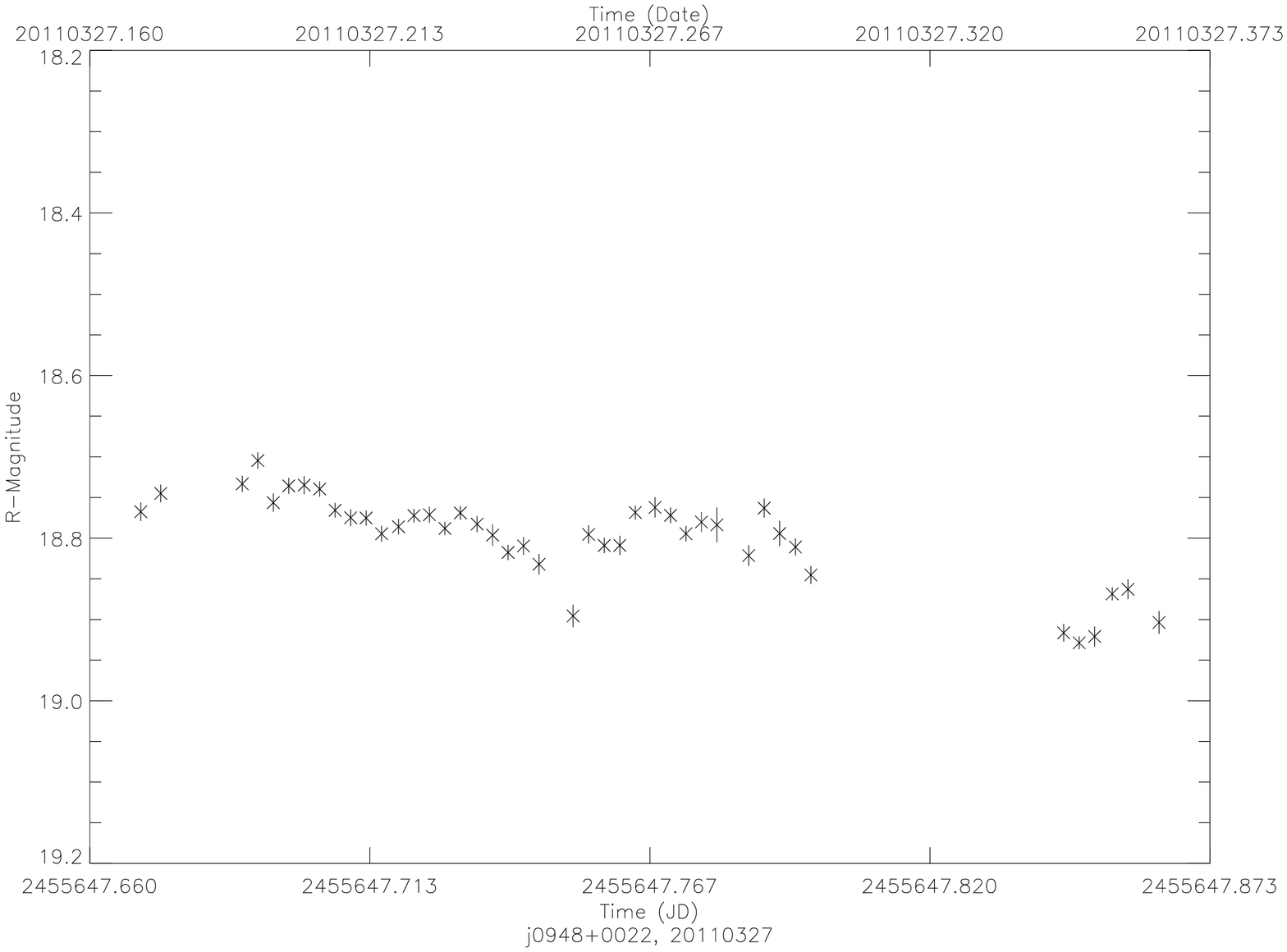}\includegraphics[scale=0.3,angle=90]{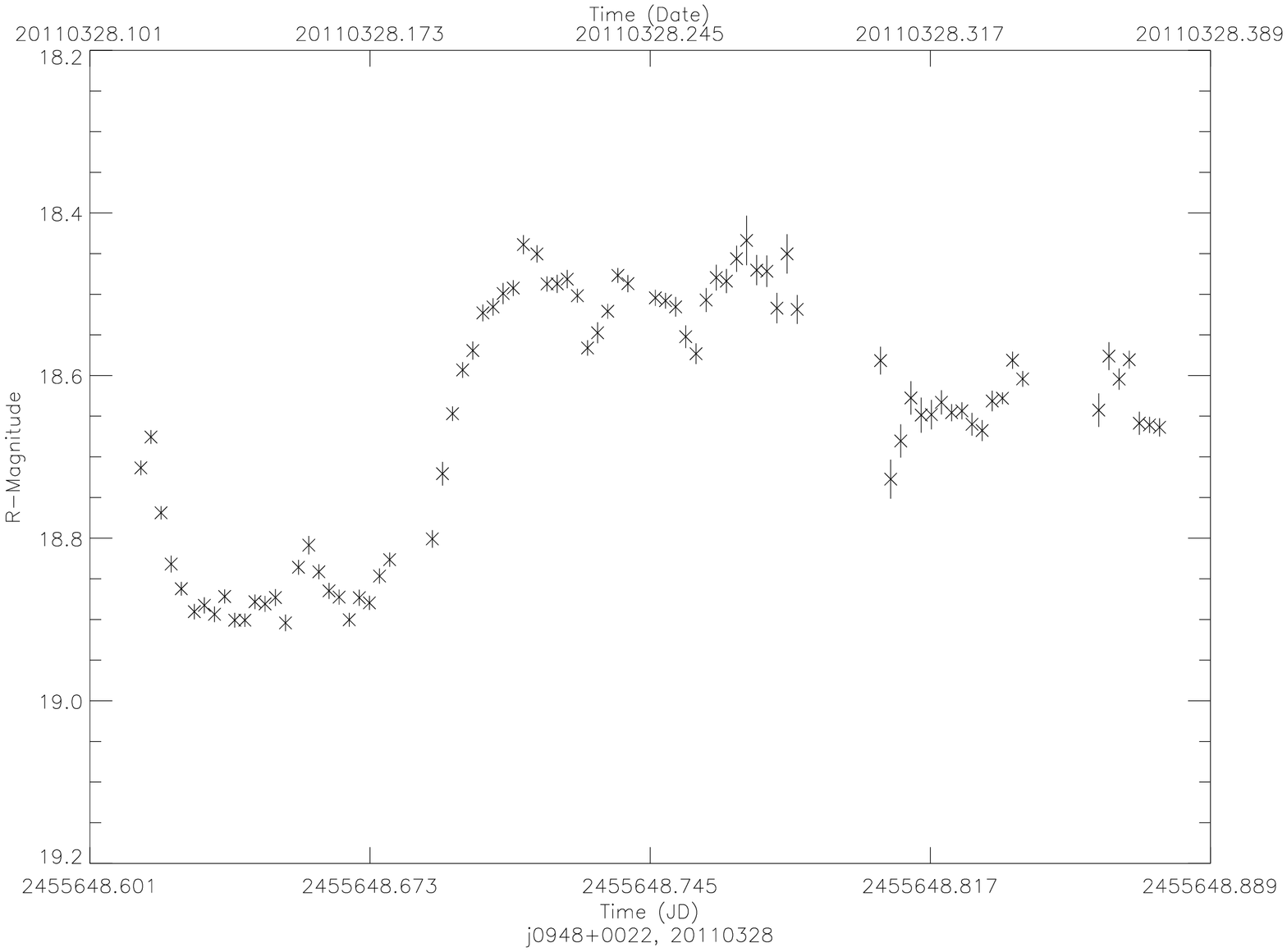}}

\noindent \centerline{\includegraphics[scale=0.3,angle=90,origin=c]{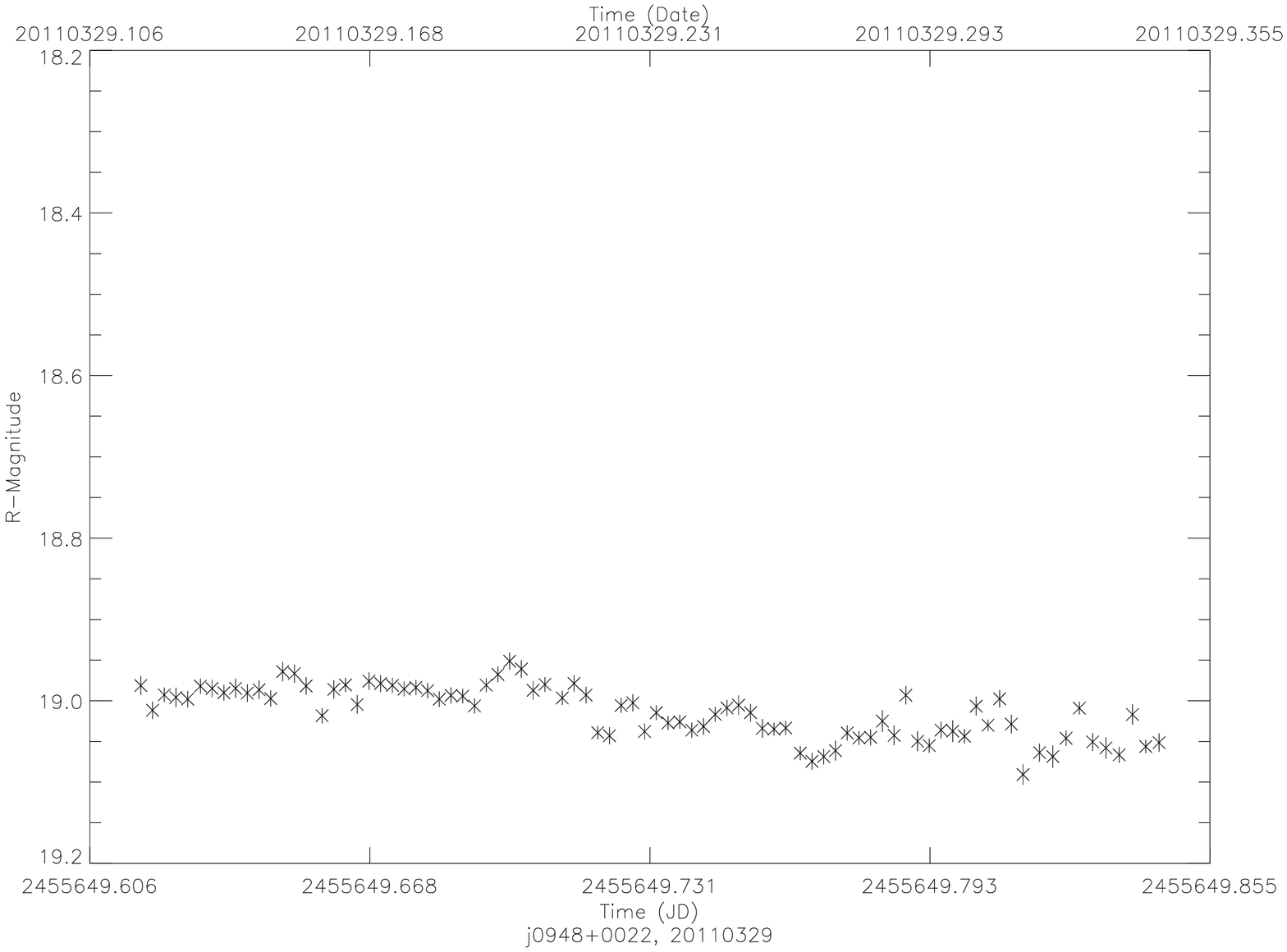}\includegraphics[scale=0.3,angle=90,origin=c]{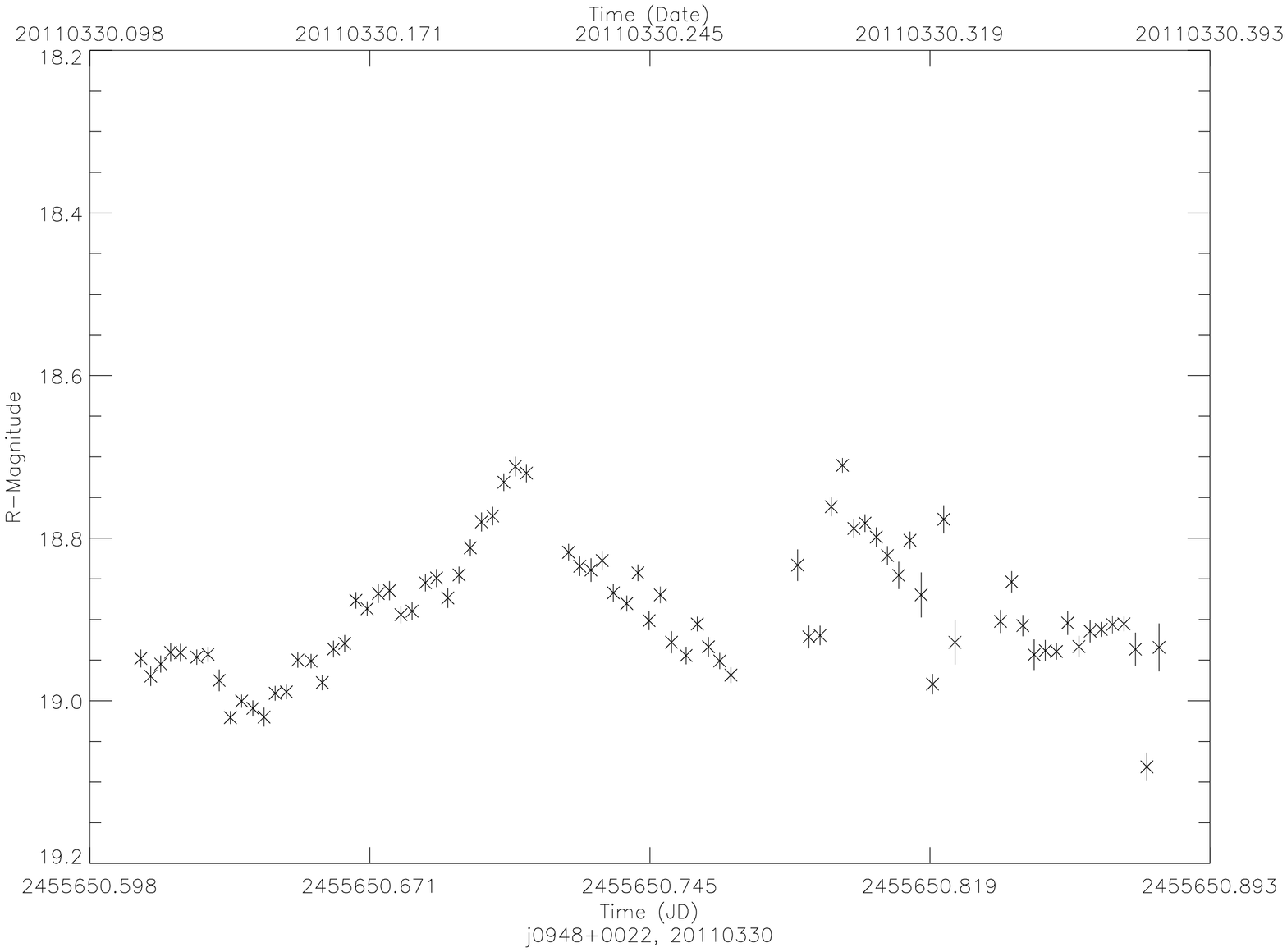}}

\noindent \centerline{\includegraphics[clip,scale=0.3,angle=90,origin=c]{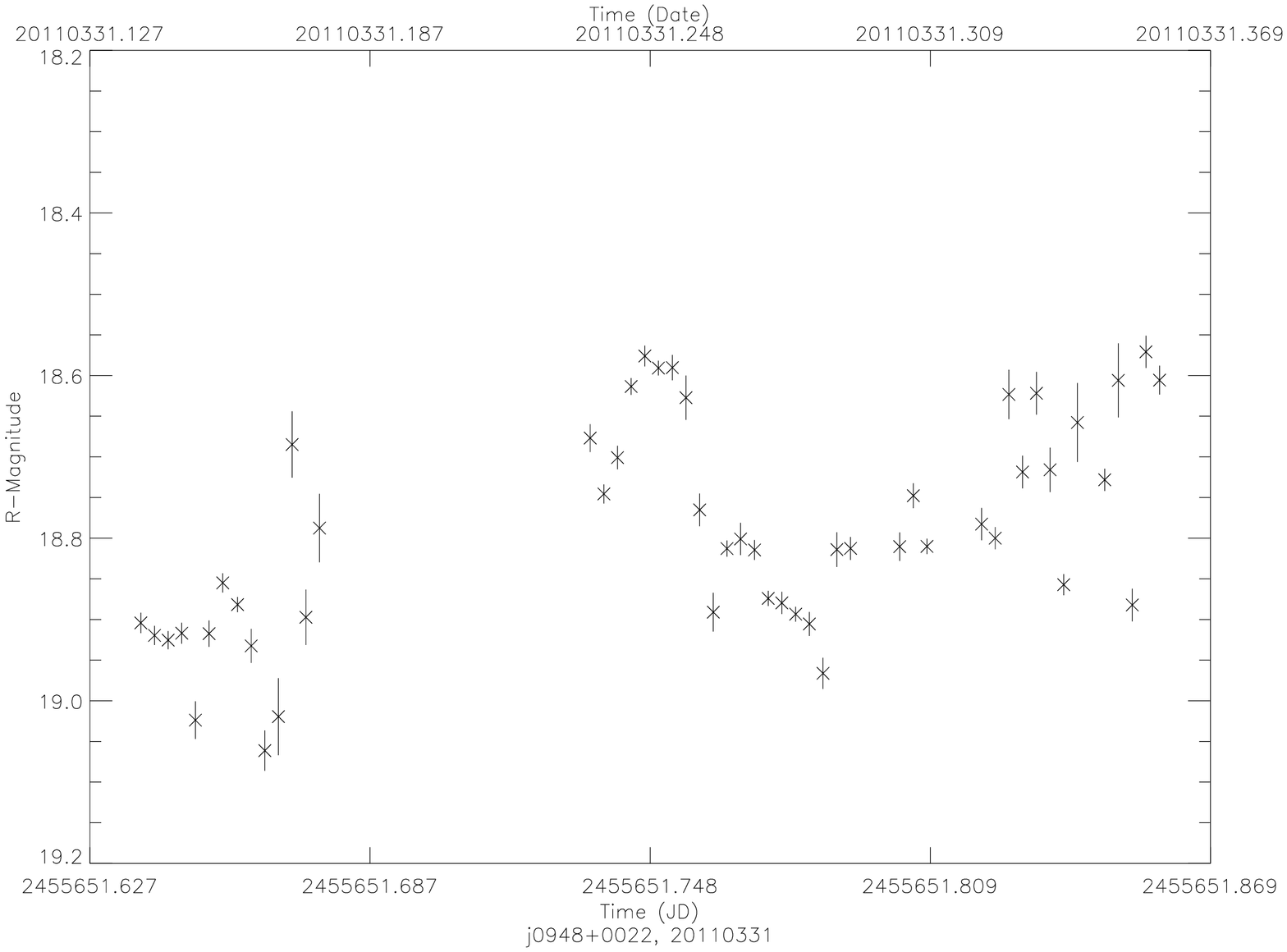}\includegraphics[clip,angle=90,origin=c,scale=0.3]{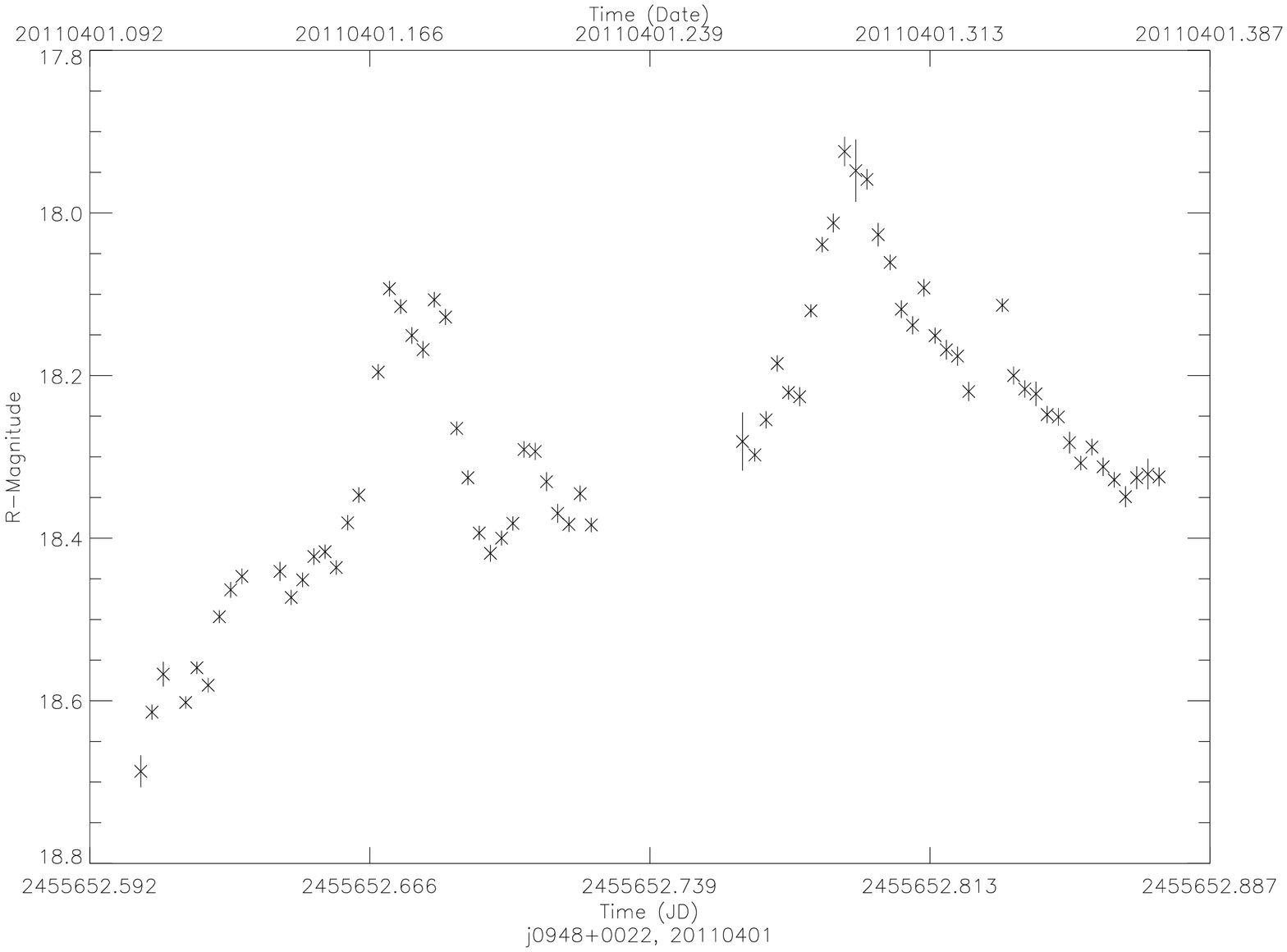}}\caption{\selectlanguage{american}%
Microvariability\foreignlanguage{english}{ behavior of J0948+0022
across the same six nights as the previous figure. All light curves
are plotted on a one-magnitude scale, though that of the final night
(April 1\textsuperscript{st}) is displaced slightly from the previous
five. Calendar dates are given in yyyymmdd.xxx format, where xxx is
the fractional day. \label{fig:hourly_lc}}\selectlanguage{english}%
}
\end{figure*}

As mentioned in section 2 of this paper, the LSW has also published
a finding chart for the field of J0948+0022%
\footnote{The finding chart in question can be found at: http://www.lsw.uni-heidelberg.de/projects/extragalactic/charts/0948+002.html%
}. However, the LSW chart lists different magnitudes for the stars
also appearing in Figure \ref{fig:Finding_Chart} of this paper; attempts
to determine the exact methodology used to derive these values were
unsuccessful. Offsets of 1.25 to 1.50 magnitudes in the R band exist
between the two charts. Normally, it might be assumed that the stars
in question were poorly chosen variable sources. However, as detailed
in section 2 these stars are known to be stable to within the uncertainty
on time scales at least as long as one year. Therefore, an attempt
to reconcile the differences between the charts had to be made least
the results of this paper be called into question. 

Normally, the magnitudes of standard stars are found using all-sky
photometry. In the analysis presented in this paper, however, these
magnitudes were derived using differential photometry and an equation
of the form 
\[
1:\;(m_{R,object}-m_{R,Landolt})_{observed}=(m_{R,object}-m_{R,Landolt})_{apparent}+k_{R}(X_{object}-X_{Landolt})
\]
where m\textsubscript{R, object}denotes both the observed and the
apparent magnitudes of a chosen object, m\textsubscript{R, Landolt}denotes
the observed and apparent magnitudes of a star taken from the Landolt
equatorial list, k is the extinction coefficient in the R band, and
X is the airmass. The extinction coefficient at Lowell observatory
during the time of observations was found to equal 0.03, with a standard
deviation of 0.01. In turn, the J0948+0022 and Landolt fields were
observed so that (X\textsubscript{object} - X\textsubscript{Landolt})
was always less than 0.02. This implies that the k\textsubscript{R}(X\textsubscript{object}
- X\textsubscript{Landolt}) term in equation (1) will always be on
the order of 10\textsuperscript{-4}. As the observational uncertainties
are themselves on the order of 10\textsuperscript{-2}, this term
is negligible and can therefore be ignored, allowing for a straightforward
calculation of the true apparent magnitude of the object once the
other quantities are known.

To further check the accuracy of this methodology, an attempt was
made to re-derive the magnitudes of the standard stars appearing in
two blazar fields the LSW has posted on their website. The same Landolt
stars used on the J0948+0022 objects were used again, and the Hall
telescope had acquired all of the relevant images. This gave a total
of 14 test objects. Again, the PEGA values and the LSW values were
not in agreement, though this time the offsets were much smaller,
ranging from 0.20 to 0.38 magnitudes. However, the blazar images had
been observed three weeks prior to the observations of the Landolt
stars, whereas the J0948+0022 field objects had been observed on the
same night. Also unlike the Landolt fields, the blazar images were
taken on a non-photometric night, and had been chosen for this analysis
purely because they were the only other available fields observed
within a month of the Landolt images that the LSW also had finding
charts for. It is therefore likely that the relatively minor offsets
between the two sources are due to a combination of atmospheric and
instrumental effects, rather than a true disagreement on the intrinsic
magnitudes of these objects.

As a further check, and as a means of avoiding the issues raised in
the above analysis, an attempt to re-derive the magnitudes of the
four standard stars the LSW lists for the field of Bl Lac was made%
\footnote{The Bl Lac chart can be found at: http://www.lsw.uni-heidelberg.de/projects/extragalactic/charts/2200+420.html%
}. In this instance, instead of using the Landolt stars in the calculation,
the magnitudes of each of the four Bl Lac stars were derived using
the other three. Since both the object and calibrating stars were
in the same image, any atmospheric or instrumental effects could be
avoided. Using this method, exactly the same magnitudes (to within
the uncertainty) were found as are listed on the LSW finding chart.
This indicates that the methodology used within this paper is not
the ultimate source of the disagreement for the objects in the field
of J0948+0022.

Finally, a random image of J0948+0022 was chosen and a final analysis
was performed on the stars selected by each group. First, in a method
similar to the above, the magnitudes of the nine stars appearing in
Figure \ref{fig:Finding_Chart} were calculated by treating each as
an unknown object and using the other eight as standards. The PEGA
values were assumed to be correct for each star's true apparent magnitude.
All nine objects were found to have the expected magnitudes, with
the highest deviation being 0.05 magnitudes for star 9 (star C on
the LSW list). This value is just above the associated uncertainty
for the object.

The same analysis was then performed on the three stars appearing
on the LSW chart using the magnitudes they claim for each object.
In this instance, much higher deviations from the expected values
were observed, with star C (number 9 in the PEGA list) again being
the worst offender with a deviation of 0.18 magnitudes. Therefore,
unlike those listed in this paper, it appears that the LSW values
for these stars are not internally consistent. 

\clearpage{}

\begin{table}
\centering{}%
\begin{tabular}{|c|c|c|c|c|}
\hline 
Telescope  & Observatory  & Detector  & Scale (arcsec/pix)  & Field of View\tabularnewline
\hline 
Perkins  & Lowell  & PRISM  & 0.780 (2x2 binning)  & 13.3' x 13.3'\tabularnewline
\hline 
Hall  & Lowell  & NASA42  & 0.740 (2x2 binning)  & 25.3' x 25.3'\tabularnewline
\hline 
NURO  & Lowell  & NASAcam  & 0.456  & 15.6' x 15.6'\tabularnewline
\hline 
1.3 meter  & CTIO  & ANDICAM  & 0.371  & 6' x 6'\tabularnewline
\hline 
\end{tabular}\caption{Instrument Specifications\label{tab:Inst-Specs}}
\end{table}

\begin{table*}[b]
\noindent \centering{}%
\begin{tabular}{|c|c|c|c|c|c|c|c|c|c|}
\hline 
Star Number  & 1  & 2  & 3  & 4  & 5  & 6  & 7  & 8  & 9\tabularnewline
\hline 
Magnitude  & 18.64  & 17.62  & 17.84  & 17.11  & 18.53  & 17.77  & 18.46  & 17.74  & 17.74\tabularnewline
\hline 
Uncertainty  & 0.04  & 0.02  & 0.02  & 0.03  & 0.05  & 0.04  & 0.02  & 0.03  & 0.04\tabularnewline
\hline 
\end{tabular}\caption{Check star information for Figure 1.\label{tab:Check_Star_Info}}
\end{table*}

\end{document}